\documentclass{IEEEcsmag}

\usepackage[colorlinks,urlcolor=blue,linkcolor=blue,citecolor=blue]{hyperref}
\expandafter\def\expandafter\UrlBreaks\expandafter{\UrlBreaks\do\/\do\*\do\-\do\~\do\'\do\"\do\-}
\usepackage{upmath,color}
\usepackage[super]{natbib}

\jvol{XX}
\jnum{XX}
\paper{8}
\jmonth{Month}
\jname{IEEE Intelligent Systems}
\jtitle{Publication Title}
\pubyear{2024}

\begin{document}

\sptitle{AI4Education}

\title{Foundation Models for Education: Promises and Prospects}

\author{Tianlong Xu}
\affil{Squirrel Ai Learning, Bellevue, USA}

\author{Richard Tong}
\affil{Squirrel Ai Learning and IEEE Artificial Intelligence Standards Committee, USA}

\author{Jing Liang}
\affil{Squirrel Ai Learning, Shanghai, China}

\author{Xing Fan}
\affil{Squirrel Ai Learning, Shanghai, China}

\author{Haoyang Li}
\affil{Squirrel Ai Learning, Shanghai, China}

\author{Qingsong Wen}
\affil{Squirrel Ai Learning, Bellevue, USA}

\markboth{THEME/FEATURE/DEPARTMENT}{THEME/FEATURE/DEPARTMENT}

\begin{abstract}\looseness-1 With the advent of foundation models like ChatGPT, educators are excited about the transformative role that AI might play in propelling the next education revolution. The developing speed and the profound impact of foundation models in various industries force us to think deeply about the changes they will make to education, a domain that is critically important for the future of humans. 
In this paper, we discuss the strengths of foundation models, such as personalized learning, education inequality, and reasoning capabilities, as well as the development of agent architecture tailored for education, which integrates AI agents with pedagogical frameworks to create adaptive learning environments. Furthermore, we highlight the risks and opportunities of AI overreliance and creativity. Lastly, we envision a future where foundation models in education harmonize human and AI capabilities, fostering a dynamic, inclusive, and adaptive educational ecosystem. 

\end{abstract}

\maketitle

\chapteri{W}ith the emergence of foundation models and generative AI (GenAI) \cite{epstein2023art}, the implications for various science and technological domains have been rapidly explored to address real-world problems~\cite{jin2023large,pan2024integrating}. Similarly, the integration of foundation models with education has naturally arisen as a promising avenue~\cite{wang2024large,li2024bringing}, particularly as large language models (LLMs)~\cite{zhao2023survey} are inherently instructive and can function like an extremely knowledgeable ``teacher''. 
Along with the trend, it is crucial to consider how to frame foundation models for education, leverage GenAI's unique advantages, and manage potential risks to traditional education. 
Foundation models, exemplified by ChatGPT, present a dual-edged sword in education, sparking debates over academic integrity versus innovative teaching aids.
They offer the potential to assist students in studying and learning. However, concerns arise regarding the rigor of foundation models, as some early applications have shown limitations~\cite{huang2023survey}. Therefore, finding a balance between utilizing AI's benefits and maintaining academic rigor is essential for the future of education~\cite{Economist2024AI}

Our major contributions are highlighting the strengths of foundation models in personalized learning, educational inequality, and reasoning capabilities, proposing an agent architecture for education, and at the same time, warning some risks of AI in education. Our major novelty is to establish a framework of future education foundational models, strengthening on educational penalization, being prepared for the overwhelming advancement of AI capabilities, and maintaining absolute human competitiveness in key capacities such as problem-solving, critical thinking, and creativity.



\section{Strengths of Foundation Models for Education}
\subsection{Personalized Learning}
The technical prowess of foundation models is revolutionizing education, as seen in the recent advancements in LLMs \cite{denny2024generative}. These advancements enhance the capabilities of LLMs to offer tailored feedback that considers students’ personal preferences and historical interactions, dramatically improving individual learning experiences. In practice, there are already some pioneers actively pursuing such directions. 
For instance, 
Khanmigo \cite{KhanAcademy2024KhanLabs} utilizes LLMs to simulate the benefits of personal tutoring, functioning as a virtual writing coach that promotes critical thinking and problem-solving. 
Squirrel AI~\cite{SquirrelAI2024Products} develops the large adaptive model encompassing foundation models, advanced RAG, and educational AI Agent, which can capture the intricate relationships between knowledge points, topics, and students' learning abilities for better personalized learning solutions.
Duolingo's Duolingo Max \cite{Duolingo2024Max} leverages LLMs for adaptive learning through roleplay, engaging users in lifelike conversations that seamlessly integrate into their learning paths. 
These platforms demonstrate how generative AI's nuanced understanding of context and personalized interaction can enhance education, making learning more responsive and interactive, much like a human tutor's guidance. 
Furthermore, industrial advancements underscore the need to build foundation models for education, which align with the principle of fostering individuals' holistic development and nurturing talents capable of innovation and independent thought in their fields.



\subsection{Addressing Education Inequality}
One widely recognized pain point for traditional education is educational inequity, whose root causes include resource allocation, teacher training, curriculum relevance, and social factors~\cite{holmes2022ethics}. The introduction of foundation models can address these issues directly and also contribute to a broader strategy for achieving education equity. By leveraging its ability to process extensive data, foundation models can pinpoint the exact needs of diverse communities, ensuring resources are allocated more fairly. It enables personalized teacher training, breaking down geographical barriers and uplifting educational quality across all regions. Additionally, adaptive learning technologies based on foundation models can tailor educational content to individual student backgrounds, making learning more engaging and accessible. This approach not only tackles the practical aspects of educational inequity but also combats social prejudices, fostering a more inclusive and equitable educational landscape.

\subsection{Reasoning Capabilities}
With the giant parameter space established during pretraining, LLMs have developed strong reasoning capabilities that continue to grow. In this domain, leveraging foundation models and adaptive learning techniques for math education is one of the most widely explored directions~\cite{li2024automate,liang2023let}. For example, many recent LLM-based works \cite{wang2023mathcoder, zhao2023automatic} have been tested or developed as the solver to the K12 level math problems, including arithmetic, geometry, equation sets, and their performance over some math word problem datasets like GSM8K are satisfactory. To further adopt LLMs for pedagogical purposes, the follow-up study explores the research questions on whether LLMs can correct students' wrong answers~\cite{yen2023three}. 
The step-by-step reasoning capabilities of Gemini \cite{Google2024GeminiAI} and many others have shown GenAI’s strong potential in conquering sophisticated problems and positioning the “mistake steps” students might have in subjects including but actually will not be limited to mathematics. Such capabilities will be a strong addition to the teaching forces, which, in the one-on-one tutoring manner, significantly boost education effectiveness. Therefore, they need to be embedded as a strong backbone while creating the foundation models for education.

\section{Agent Architecture for Education}

\begin{figure}[h]
\includegraphics[width=8cm, height=6cm]{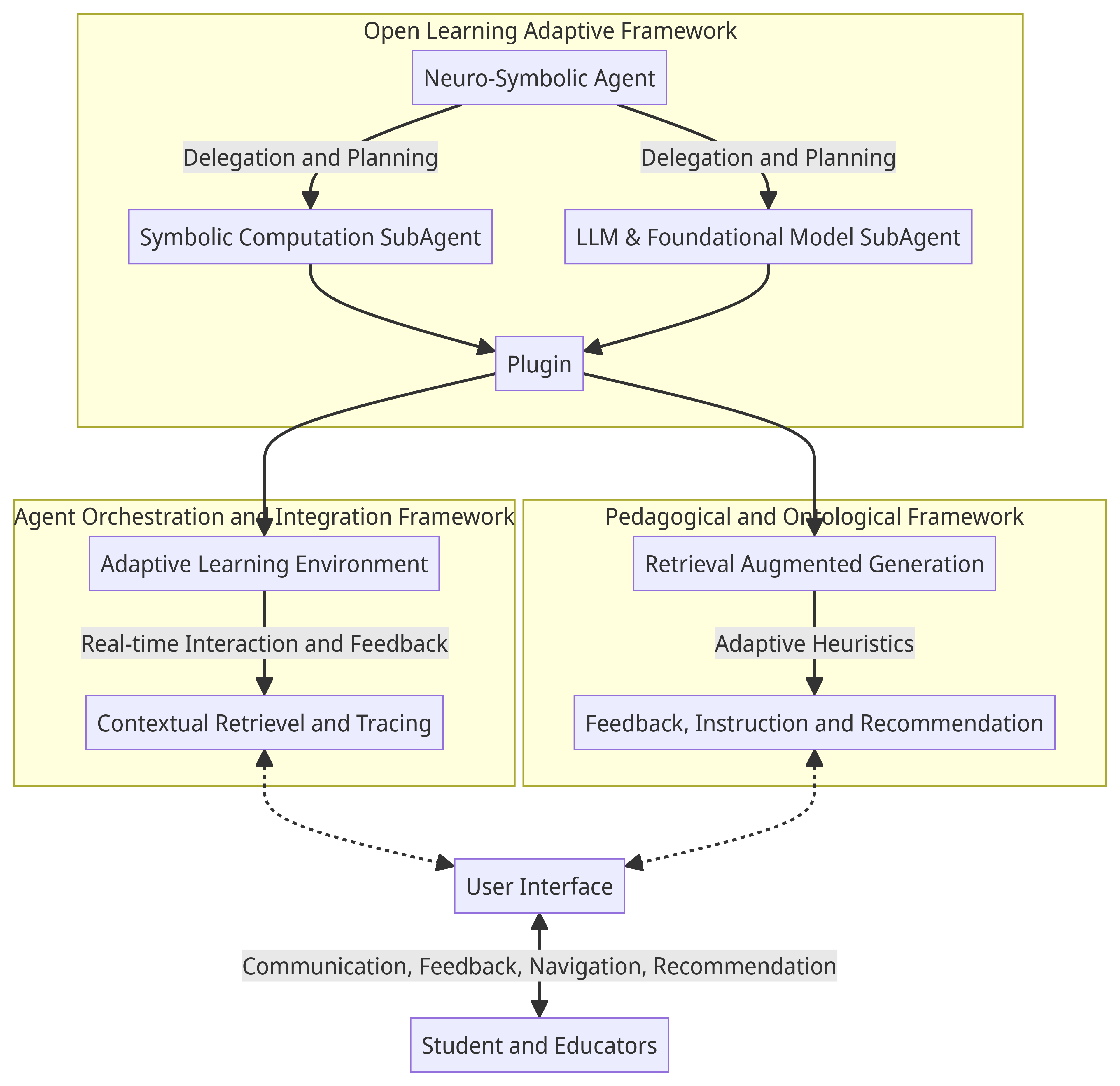}
\caption{The Agent Architecture Abstraction for Education.}
\label{fig_agent_edu}
\end{figure}

To harness the potential of foundation models in the adaptive instructional environment, we foresee a new type of system architecture built on top of AI agents, as shown in Figure~\ref{fig_agent_edu}. This architecture can manage diverse and complex inputs for various pedagogical situations, adapt to changing contexts and curricula in a self-improvable manner \cite{Tong2020Architecture}, and navigate the often ambiguous and interactive demands of students and educators\cite{TongLee2023}. It can be broken down into three components.
1) Core Agent Architecture: 
at the heart of the system are specialized agents responsible for distinct cognitive functions. These may range from diagnosis, forecasting, and problem-solving to providing psychological support. Each agent typically integrates both symbolic reasoning and neural network capabilities leveraging LLM and other foundation models. 
2) {Agent Orchestration and Integration Framework}:
this layer serves as the 'environment' that hosts the agents and enables their interaction, not only among themselves but also with external tools and platforms. A well-designed environment facilitates channel-based communication, where student behaviors, interactions, and other resources are funneled into a single session. This consolidated session offers real-time interaction and feedback mechanisms between agents, students, and educators. It also logs these activities for knowledge tracing, model refinement, or compliance monitoring. 
 3) {Pedagogical and Ontological Framework}:
beyond what is traditionally called a content management system, this framework operates at the intersection of content, learning objectives, and pedagogical strategies. It would likely be built upon an ontology framework that interlinks resources with learning goals and pedagogical heuristics. 
\vspace*{6pt}
	






\section{Risks and Potential Opportunities}

\subsection{Overreliance}
Responsible AI has been a widely discussed topic since the introduction of AI. According to the latest framework Microsoft has proposed, there are six critical components: fairness, inclusiveness, reliability \& safety, privacy \& security, transparency, and accountability \cite{Microsoft2024ResponsibleAI}. Beyond these, the potential tendency of overreliance as we continue to interact more with GenAI is worth being discussed as well. This issue mainly come from the concerns of AI implications on education, however, it could also extend to a wider range of impact given the “instructive” characteristics of most GenAI products. GenAI’s ability to provide instant information might lead to a dependency that undermines critical thinking and the motivation for self-led learning. To combat this, AI should be used to encourage deeper inquiry, not just quick answers. Integrative teaching strategies that demand independent research and critical thinking are key to preventing overreliance on AI, ensuring students retain their ability to learn autonomously. This balance is crucial for the responsible incorporation of AI in education, presenting a challenge for educators and policymakers to foster independent, inquisitive learners while leveraging AI's advantages.

\subsection{AI Creativity}
A further step in thinking beyond the overreliance is whether foundation models can be truly innovative hinges on their learning capacity. Some argue that AI systems like GPT-4 learn from vast datasets and may exhibit creativity~\cite{wang2024ai}, a view that could be implicitly supported by the "larger is better" hypothesis. However, the speculation remains open, as the extent to which AI can originate novel ideas is still unproven. Regardless of AI's potential for creativity, the emphasis in education should remain on nurturing human ingenuity. If AI is indeed capable of innovation, the challenge is to ensure it complements rather than competes with human creativity. By promoting educational frameworks that prioritize independent and critical thinking skills, we can ensure the dominance of human creativity, either using GenAI as a tool to enhance and amplify (rather than replace) the creative process, or using it as a virtuous competitor to maintain human competitiveness.

\section{Conclusion and Future Vision}

In the evolving landscape of foundation models, personalized learning emerges as a pivotal force in enriching educational experiences. It caters to the diverse needs, preferences, and abilities of each learner, thereby advancing educational equity. While technology serves to enhance these experiences, it is imperative that the essence and ultimate responsibility of decision-making remain firmly anchored in human hands. Looking forward, the envisioned educational foundation model heralds a shift towards a more dynamic, inclusive, and adaptive framework. This framework seeks to harmonize the strengths of human educators with the capabilities of GenAI technology, thereby preparing learners to navigate both foreseeable and unforeseen challenges with resilience and adaptability.

The future of education is envisioned as a realm where foundation models serve to amplify the value from human's potential and the vast amount of knowledge accumulated rather than a total replacement, such that AI and human’s coevolution can progress towards an ideal direction. Future education is tailored to every individual's unique journey, empowering each learner to excel and realize full potential. This vision presents a holistic blueprint for cultivating educational environments that elevate human capacities, ensuring that learners from all backgrounds can thrive and maintain their utmost competitiveness in problem-solving capabilities, critical thinking, and creativity. 

\def\refname{REFERENCES}
\bibliographystyle{unsrt}
\bibliography{bib}

\end{document}